\documentclass[]{aa}     
\usepackage{graphics}
\def\cc{\,{\rm cm^{-3}}}
\def\cm2{\,{\rm cm^{-2}}}
\def\kms{\,{\rm {km\,s^{-1}}}}
\def\kkms{\,{\rm {K\,km s^{-1}}}}
\def\co{\,{\rm ^{12}CO}}
\def\13co{\,{\rm ^{13}CO}}
\def\h2{\,{\rm H_{2}}}
\def\Msun{\rm M_{\odot}}

\def\aua{{\rm A\&A} }

\def\apj{{\rm ApJ} }

\def\apjs{{\rm ApJS} }
\def\apjl{{\rm ApJL} }
\def\araa{{\rm ARAA} }
\def\mnras{{\rm MNRAS} }
\def\pasj{{\rm PASJ} }

\begin{document}
 
\title{CI and CO in the center of M~51}
 
   \subtitle{}
 
\author{F.P. Israel
          \inst{1}
        R.P.J. Tilanus
          \inst{2},
          \and 
        F. Baas
          \inst{1, 2, \dagger}
} 

\offprints{F.P. Israel} 
 
\institute{Sterrewacht Leiden, P.O. Box 9513, 2300 RA Leiden, the Netherlands 
\and Joint Astronomy Centre, 660 N. A'ohoku Pl., Hilo, Hawaii, 96720, USA
}

\date{
Received ????; accepted ????}
 
\abstract{ We present $J$=2--1, $J$=3--2, $J$=4--3 $\co$ maps as well
as $J$=2--1, $J$=3--2 $\13co$ and 492 GHz [CI] measurements of the
central region in M~51. The distribution of CO is strongly
concentrated towards the spiral arms. The center itself is poor in,
though not devoid of, CO emission. The observed line intensities
require modelling with a multi-component molecular gas. A dense
component must be present ($n(\h2) \approx 10^{3}$) with kinetic
temperature $T_{\rm kin} \approx$ 100 K, combined with either a less
dense ($\approx 10^{2} \cc$) component of the same temperature, or a
more dense ($n(\h2) \approx 3 \times 10^{3} \cc$) and much cooler
($T_{\rm kin}$ = 10--30 K) component. Atomic carbon amounts are
between 5 and 10 times those of CO. Much of the molecular gas mass is
associated with the hot PDR phase. The center of M~51 has a face-on
gas mass density of about $40\pm20 \Msun$\, pc$^{-2}$, and a
well-established CO-to-H$_{2}$ conversion ratio $X$ four to five times
lower than the standard Galactic value.  \keywords{Galaxies --
individual (M~51) -- ISM -- centers; Radio lines -- galaxies; ISM --
molecules} }

\maketitle
 
\section{Introduction}

\begin{table}
\caption[]{\centerline{Galaxy parameters}}
\begin{center}
\begin{tabular}{lll}
\hline
\noalign{\smallskip}
			    & M~51 \\
\noalign{\smallskip}
\hline
\noalign{\smallskip}
Type$^{a}$     	            & SA(s)bcp \\
Radio Center :		    & \\
R.A. (B1950)$^{b}$ 	    & 13$^{h}$27$^{m}$46.3$^{s}$ \\
Decl.(B1950)$^{b}$          & +47$^{\circ}$27$'$10$''$   \\
R.A. (J2000)$^{b}$          & 13$^{h}$29$^{m}$52.7$^{s}$ \\
Decl.(J2000)$^{b}$          & +47$^{\circ}$11$'$42$''$   \\
$V_{\rm LSR}^{c,}$    	    & +464$\kms$                 \\
Inclination $i^{c}$ 	    & 20$^{\circ}$               \\
Position angle $P^{c}$      & 170$^{\circ}$ 		 \\
Distance $D^{d}$            & 9.7 Mpc                    \\
Scale           	    & 21.3$''$/kpc               \\
\noalign{\smallskip}
\hline
\end{tabular}
\label{m51parms}
\end{center}
Notes to Table 1: 
$^{a}$ RSA (Sandage $\&$ Tammann 1987);
$^{b}$ Turner $\&$ Ho (1994); 
$^{c}$ Tully (1974)
$^{d}$ Sandage \& Tamann (1975)
\end{table}

\begin{table*}
\caption[]{\centerline{Observations Log}}
\begin{center}
\begin{tabular}{rccccccccc}
\hline
\noalign{\smallskip}
Transition & Date    	& Freq	& $T_{\rm sys}$ & Beam 	& $\eta _{\rm mb}$ & t(int) & \multicolumn{3}{c}{Map Parameters} \\
& & & & Size & & & Points & Size & Spacing \\
	   & (MM/YY) 	& (GHz)	& (K)	  & ($\arcsec$) & 	       	   & (sec) & & ($\arcsec$) & ($\arcsec$) \\
\noalign{\smallskip}
\hline
\noalign{\smallskip}
$\co$ $J$=2--1	  & 12/90; 05/91 & 230 &  920 & 21 & 0.69 &  480 &  40 & 30$\times$50 & 10 \\
$\co$ $J$=3--2    & 12/93        & 345 & 1475 & 14 & 0.53 &  400 &  18 & 40$\times$50 & 12 \\
	   	  & 05/97; 05/01 &     &  660 &    & 0.60 &   70 & 400 & 60$\times$120 & 6 \\
$\co$ $J$=4--3    & 03/94; 04/96 & 461 & 3950 & 11 & 0.50 &  600 &  16 & 24$\times$20 &  6 \\
		  & 12/01        &     & 2100 &	   & 0.52 & 1200 &   6 & 18$\times$12 &  6 \\
\noalign{\smallskip}
\hline
\noalign{\smallskip}
$\13co$ $J$=2--1 & 06-95         & 220 &  370 & 21 & 0.69 & 2520 &  1 & ---  & ---  \\
$\13co$ $J$=3--2 & 05-97; 05/01  & 330 &  600 & 14 & 0.63 & 3600 &  9 & 15$\times$15 & 7.5 \\
\noalign{\smallskip}
\hline
\noalign{\smallskip}
CI $^{3}$P$_{1}$--$^{3}$P$_{0}$ & 11-94 & 492 & 6675 & 10 & 0.53 &  960 & 4 & --- & --- \\
\noalign{\smallskip}
\hline
\end{tabular}
\label{m51log}
\end{center}
\end{table*}

Molecular gas is a major constituent of the interstellar medium in 
galaxies. This is particularly true for star-forming complexes in the 
spiral arms, but strong circumnuclear concentrations of molecular gas 
are frequently found also in the inner kiloparsec of spiral galaxies.
We have observed a sample of nearby spiral galaxies in various CO 
transitions and in the 492 GHz $^{3}$P$_{1}$--$^{3}$P$_{0}$ [CI] 
transition in order to determine the physical condition of molecular 
gas in their inner parts. Results for NGC~253 (Israel et al. 1995), 
NGC~7331 (Israel $\&$ Baas 1999), NGC~6946, M~83 = NGC~5236 
(Israel $\&$ Baas 2001) as well as IC~342 and Maffei 2 (Israel $\&$ 
Baas 2003) have been published. In this paper, we present results 
obtained for the interacting, two-armed spiral M~51 = NGC~5194 (see
Table\,\ref{m51parms}). 

M~51 was one of the first galaxies mapped in CO line emission, and it
has been observed many times since (Rickard et al. 1977; Scoville $\&$ 
Young 1983; Rydbeck et al. 1985; Sandqvist et al. 1989; Garcia-Burillo 
et al. 1993a, b; Berkhuijsen et al. 1993; Nakai et al. 1994; Young et 
al. 1995; Mauersberger et al. 1999; Wielebinski et al. 1999; Dumke et 
al. 2001; Paglione et al. 2002; Tosaki et al. 2002), including mapping 
observations with millimeter arrays (Lo et al. 1987; Rand $\&$ Kulkarni 
1990; Rand 1993; Scoville et al. 1998; Aalto et al. 1999; Sakamoto et al. 
1999; Regan et al. 2001).

\section{Observations}

\begin{figure*}
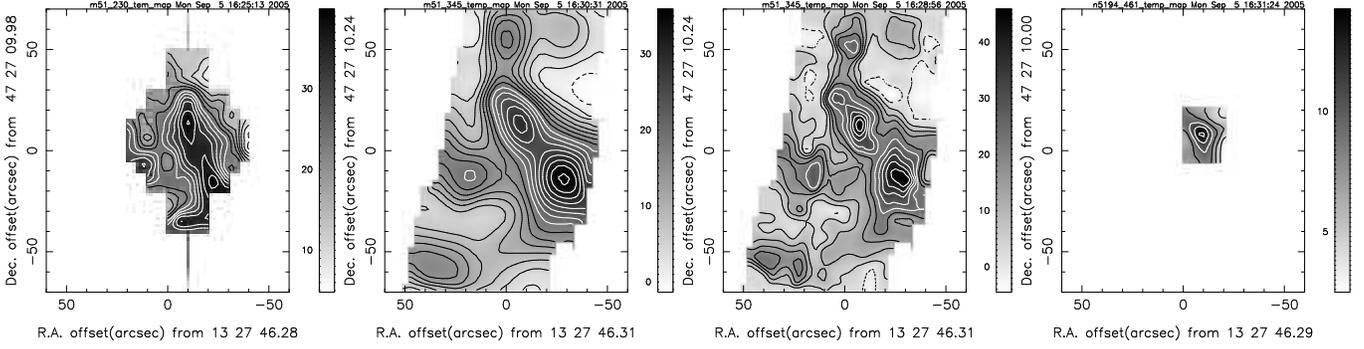

\unitlength1cm
\begin{minipage}[b]{4.4cm}
\resizebox{4.4cm}{!}{\rotatebox{270}{\includegraphics*{3096fi1a.ps}}}
\end{minipage}
\hfill
\begin{minipage}[t]{4.4cm}
\resizebox{4.4cm}{!}{\rotatebox{270}{\includegraphics*{3096fi1b.ps}}}
\end{minipage}
\hfill
\begin{minipage}[b]{4.4cm}
\resizebox{4.4cm}{!}{\rotatebox{270}{\includegraphics*{3096fi1c.ps}}}
\end{minipage}
\hfill
\begin{minipage}[b]{4.4cm}
\resizebox{4.4cm}{!}{\rotatebox{270}{\includegraphics*{3096fi1d.ps}}}
\end{minipage}
\label{m51contour}
\caption[] {Contour maps of emission from M~51 integrated over the
velocity range $V_{\rm LSR}$ = 300 to 600 $\kms$. North is at top.
Offsets are marked in arcsec relative to the M~51 nucleus at (0,0)
with B1950 coordinates given corresponding to J2000 coordinates
$\alpha_{\circ} = 13^{h}29^{m}52.^{s}7, \delta_{\circ} =
+47^{\circ}11^{'}43^{''}$ (cf. Table\,\ref{m51parms}).  Left to right:
CO $J$=2--1 at 21$''$ (985 pc) resolution, CO $J$=3--2 convolved to
21$''$, CO $J$=3-2 at 14$''$ (656 pc) resolution and CO $J$=4--3
convolved to 14$''$ resolution.  Contour values are linear in $\int
T_{\rm mb} {\rm d}V$. Contour steps are 4 $\kkms$ (2--1 full
resolution, 3--2 convolved, 4--3 convolved) and 8 $\kkms$ (3-2 full
resolution) and start at 0. }
\end{figure*}

All observations described in this paper were carried out with the 15m
James Clerk Maxwell Telescope (JCMT) on Mauna Kea (Hawaii)
\footnote{The James Clerk Maxwell Telescope is operated on a joint
basis between the United Kingdom Particle Physics and Astrophysics
Council (PPARC), the Netherlands Organisation for Scientific Research
(NWO) and the National Research Council of Canada (NRC).}. At the
epoch of the mapping observations (1990-1996) the absolute pointing of
the telescope was good to about $3''$ r.m.s. as provided by pointing
observations with the JCMT submillimetre bolometer.  The spectra were
calibrated in units of antenna temperature $T_{\rm A}^{*}$, correcting
for sideband gains, atmospheric emission in both sidebands and
telescope efficiency. Calibration was regularly checked by observation
of a standard line source. Further observational details are given in
Table\,\ref{m51log}.  Most of the observations were carried out with
the now defunct receivers A2, B3i and C2. Observations in 2001 were
obtained with the current receivers B3 (330/345 GHz) and W/C (461
GHz). Full details on these receivers can be found at the JCMT website
(http://docs.jach.hawaii.edu/JCMT/HET/GUIDE/). Note that the angular
beamsizes used correspond to linear resolutions ranging from 470 to
1000 pc at the distance assumed for M~51.  Up to 1993, we used a 2048
channel AOS backend covering a band of 500 MHz ($650\kms$ at 230
GHz). After that year, the DAS digital autocorrelator system was used
in bands of 500 and 750 MHz. Integration times (on+off) given in
Table\,\ref{m51log} are typical values appropriate to the
maps. Because of the relatively good weather conditions and its very
close sampling, the 345 GHz $J$=3--2 $^{12}$CO map (shown in
Fig.\,\ref{m51contour}) should be considered the most reliable. Both
the 230 GHz $J$=2--1 $^{12}$CO and the 461 GHz $J$=4--3 $^{12}$CO maps
were obtained under less favourable weather conditions (as evidenced
by the system temperatures listed in Table\,\ref{m51log}). Features in
the 230 GHz map show offsets from their counterparts in the 345 GHz
map by as much as $8''$, and the peak in the 461 GHz map is offset
from its 345 GHz counterpart by $5.5''$. As these map position
differences are less than three resp. two times the r.m.s. pointing
accuracy, we do not consider them to be physically significant.

M~51 is observed almost face-on, lines are narrow and sufficient free
baseline was available to subtracted second or even third order
baselines from the profiles. Finally, all spectra were scaled to a
main-beam brightness temperature, $T_{\rm mb}$ = $T_{\rm A}^{*}$/$\eta
_{\rm mb}$; values for $\eta _{\rm mb}$ are given in
Table\,\ref{m51log}.

Much of the following analysis is based on line ratios at two specific
positions A and B. Position A is that of the (radio) nucleus, position
B is centered on a large and bright CO complex (called Giant Molecular
Association 1 by Tosaki et al. 1994) in one of the inner spiral arms
at a radial distance of 0.9 kpc northwest of the nucleus. Measurements
of these positions are much more accurate than those of individual map
positions. They were usually observed more than once, with better
pointing (typically $1.5'' - 2''$ r.m.s.) and with significantly
longer integration times resulting not only in higher signal-to-noise
ratios but also in better baselines and higher position reliability.
Relatively accurate line intensities for full-resolution profiles at
the two positions are summarized in Table\,\ref{m51int}.  In this
Table, we also give extrapolated line intensities at other
resolutions, derived as follows.  We compared intensities at the two
positions in maps at full-resolution and in maps convolved to the
desired resolution.  The accurate full-resolution measurements in
Table\,\ref{m51int} were then multiplied by the empirical scaling
factors thus determined.  We verified that this is a robust procedure.
Pointing errors do result in different intensities at any given
position, but scaling factors change much more slowly.

\begin{table*}[t]
\caption[]{\centerline{Central CO and CI line intensities in M~51}}
\begin{center}
\begin{tabular}{rlrrcrc}
\hline
\noalign{\smallskip}
\multicolumn{2}{l}{Transition} & Resolution & $T_{\rm mb}$ & $\int T_{\rm mb}$d$V$ & $T_{\rm mb}$ & $\int T_{\rm mb}$d$V$ \\
&       & ($\arcsec$)    & (mK) & ($\kkms$) & (mK) & ($\kkms$) \\
\noalign{\smallskip}
\hline
\noalign{\smallskip}
& & & \multicolumn{2}{c}{Position A:} & \multicolumn{2}{c}{Position B:}\\
& & & \multicolumn{2}{c}{Center}      & \multicolumn{2}{c}{NW Arm GMA-1}\\
\noalign{\smallskip}
\hline
\noalign{\smallskip}
$\co$   & $J$=2--1 & 21 &  260 &  38.8$\pm$8.5 &  805 &  54.0$\pm$7.5 \\
        &          & 43 &  --- &  44.5$\pm$9.0 &  --- &  36.0$\pm$7.2\\
$\13co$ &	   & 21 &  --- &      ---      &   77 &   7.5$\pm$1.0 \\
$\co$   & $J$=3--2 & 14 &  375 &  42.7$\pm$6.5 & 1535 &  64.5$\pm$9.5 \\
        &	   & 21 &  --- &  30.2$\pm$6.0 &  --- &  55.0$\pm$9.5 \\
$\13co$ &          & 14 &   28 &   3.3$\pm$0.7 &   43 &   5.5$\pm$1.5 \\
$\co$   & $J$=4--3 & 11 &  160 &  27.0$\pm$4.5 &  520 &  52.5$\pm$7.5 \\
        &	   & 14 &  --- &  23.0$\pm$4.5 &  --- &  33.0$\pm$7.0 \\
        &	   & 21 &  --- &  19.0$\pm$3.8 &  --- &  24.0$\pm$4.0 \\
\noalign{\smallskip}
[CI] & $^{3}$P$_{1}$--$^{3}$P$_{0}$ & 10 & --- &  --- & 565 & 28$\pm$5 \\
	&          & 21 &  --- &      ---      &  --- & (13$\pm$4)   \\
\noalign{\smallskip}
\hline
\end{tabular}
\label{m51int}
\end{center}
Note to Table 3: Position A: beam centered on nucleus; 
Position B: beam centered on Giant Molecular Association 1 in
NW spiral arm at offsets $\Delta\alpha = -10''$, 
$\Delta\delta=+15''$ with respect to the nucleus.
\end{table*}

\begin{figure*}
\unitlength1cm
\begin{minipage}[b]{9.0cm}
\resizebox{8.8cm}{!}{\rotatebox{270}{\includegraphics*{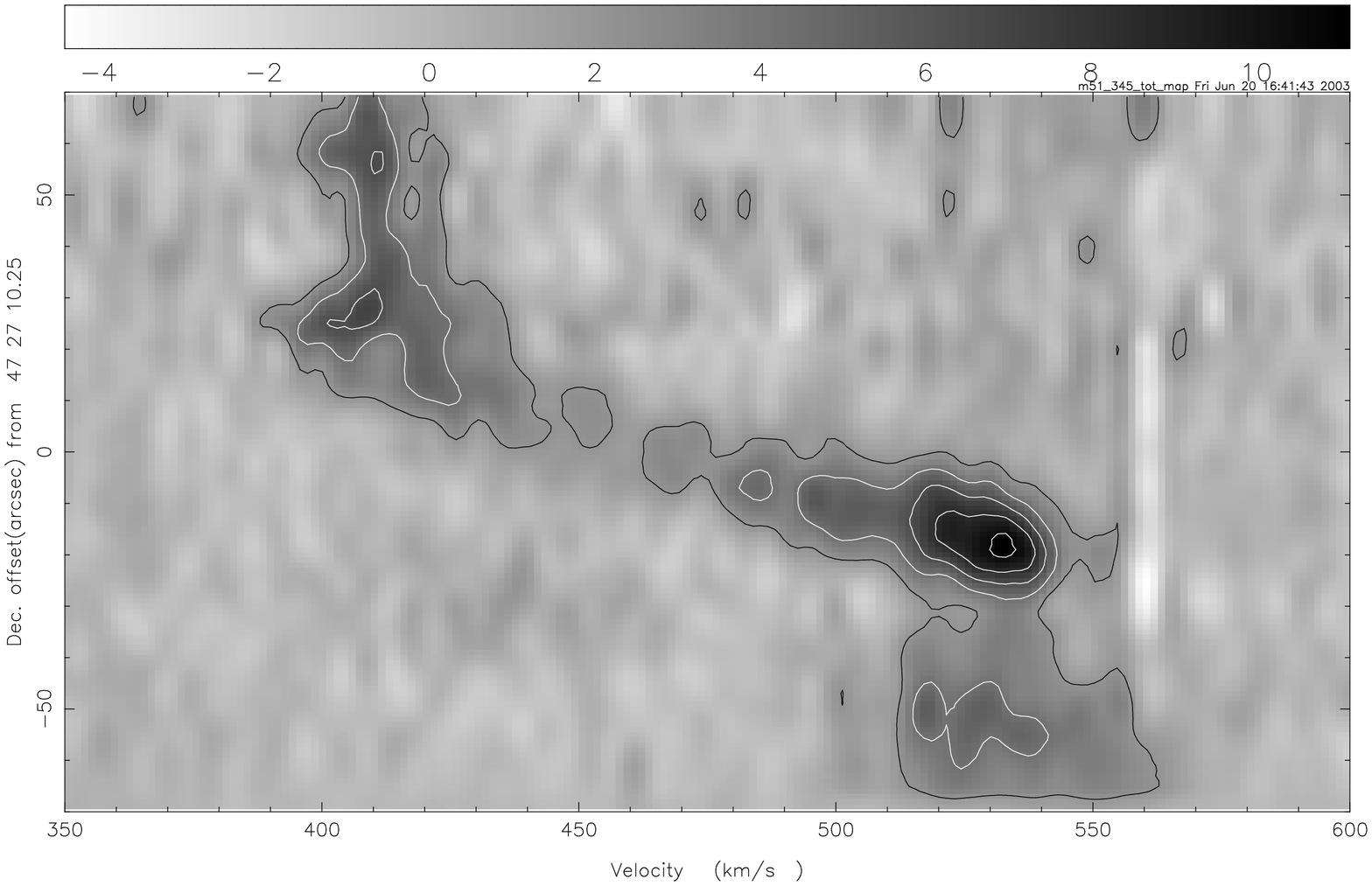}}}
\end{minipage}
\hfill
\begin{minipage}[t]{9.0cm}
\resizebox{8.8cm}{!}{\rotatebox{270}{\includegraphics*{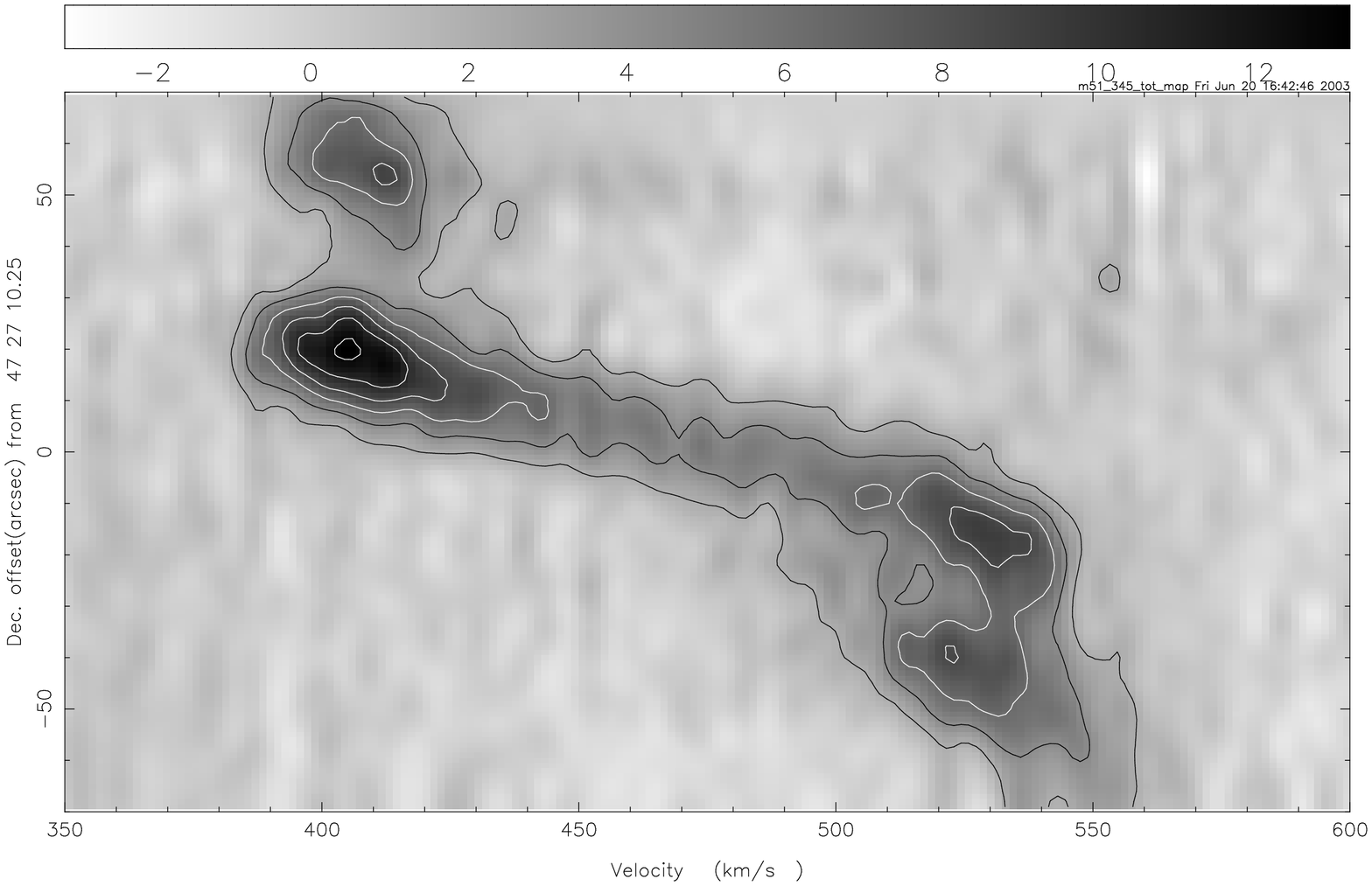}}}
\end{minipage}
\label{m51pv}
\caption[]
{Position-velocity maps of CO $J$=3--2 emission from M~51 in 
position angle 10$^{\circ}$. Left: east of the nucleus, averaged 
$\Delta \alpha$ = 20$''$--0$''$; right: west of the nucleus,
averaged over $\Delta \alpha$ = 0$''$-- -20$''$. Contour values 
are linear in steps of $T_{\rm mb}$ = 4 K, starting at step 1. 
Horizontal scale is $V_{\rm LSR}$
}
\end{figure*}

\section{Results}

\subsection{CO distribution}

The distribution of molecular gas as traced by CO shows a minimum at
the position of the galaxy nucleus (position $0'',0''$ in
Fig.\,\ref{m51contour}).  CO is strongly concentrated towards the major
spiral arms.  The CO maps in Fig.\,\ref{m51contour} clearly show the
outline of the inner parts of the western spiral arm, as well as part
of the inner eastern arm.  In all observed transitions, the emission
peaks coincide with the brightest CO peaks seen in the $J$=1--0 maps
(see e.g. Scoville et al; 1998 or Regan et al. 2001) which are
associated with star-forming regions.

The north-south position-velocity maps (Fig.\,\ref{m51pv}),
paralleling the major axis of M~51, show emission covering a velocity
range of 150 $\kms$, which becomes $440 \kms$ after correction for
inclination. The velocity distributions east and west of the major
axis are practically identical, at least at our resolution. Within a
radius $R\,<\,30''$ the molecular gas is in rapid solid-body rotation.
Again, the relative lack of molecular gas in the center (position +464
km/s, $0''$) is obvious, as is the strong molecular emission near the
radius where differential rotation takes over.  As the pV-maps of M~51
are virtually indistinguishable from the major axis pV-diagrams of
other, often more highly tilted galaxies, they illustrate an
interesting point.  In those galaxies, the existence of a compact
central feature with a steep linear velocity gradient is often
interpreted as the signature of a circular-symmetric physical
structure such as e.g. a rotating circumnuclear disk or torus.
However, the almost face-on image of M~51 clearly shows the absence of
such large-scale structure and the diagrams in Fig.\,\ref{m51pv}
simply reflect the strong molecular enhancement of the inner spiral
arms.

\subsection{Line ratios}

\begin{table*}
\begin{center}
\caption[]{\centerline{Integrated line ratios in the centre of M~51}}
\begin{tabular}{lcccc}
\hline
\noalign{\smallskip}
Transitions & \multicolumn{2}{c}{Center} &  \multicolumn{2}{c}{NW Arm GMA-1} \\ 
            & Observed & Modelled$^{a}$   & Observed & Modelled$^{a}$ \\
\noalign{\smallskip}
\hline
\noalign{\smallskip}
$\co$ (1--0)/(2--1)$^{b}$  & 1.25$\pm$0.10 & 1.28 & 1.40$\pm$0.20 & 1.35 \\
$\co$ (3--2)/(2--1)$^{c}$  & 0.78$\pm$0.23 & 0.75 & 1.02$\pm$0.16 & 0.87 \\
$\co$ (4--3)/(2--1)$^{c}$  & 0.49$\pm$0.10 & 0.41 & 0.44$\pm$0.08 & 0.46 \\
\noalign{\smallskip}
$\co$/$\13co$ (1--0)$^{d}$ &    7$\pm$1.5  &  6.7 &    8$\pm$1.5  &  8.2 \\
$\co$/$\13co$ (2--1)$^{c,d}$ &  6$\pm$1    &  6.9 &  8.7$\pm$1.2  &  8.1 \\
$\co$/$\13co$ (3--2)$^{e}$ & 12.9$\pm$1.8  & 12.2 & 12.5$\pm$1.7  & 13.2 \\
\noalign{\smallskip}
CI/CO(2--1)$^{c}$   	   & 0.22$\pm$0.06 &  --- & 0.24$\pm$0.08 & ---  \\
CII/CO(2--1)$^{f}$  	   & 0.46$\pm$0.09 &  --- &      ---      & ---  \\
\noalign{\smallskip}
\hline
\label{m51rat}
\end{tabular}
\end{center}
Notes: a. See Table\,\ref{m51model}.  b. Ratios estimated from data by
Nakai et al. 1994, NRO at 16$''$ resolution; Garcia-Burillo et
al. 1993, IRAM at 22$''$.  c. This Paper, JCMT at 21$''$ resolution.
d. Ratios estimated from data presented by Rickard $\&$ Blitz 1985,
NRAO 65$''$, Garcia-Burillo et al. 1993, IRAM at 22$''$, Matsushita et
al.  1997, NMA at 6$''$; and Paglione et al. 2001, FRAO at 45$''$
resolution.  e. This Paper; JCMT at 14$''$ resolution.  f. From
Crawford et al. (1985) and Stacey et al. (1991), KAO at 55$''$
resolution; consistent with ISO flux (Negishi et al., 2001); JCMT
convolved to 55$''$ resolution.
\end{table*}

Using our own and literature data, we have determined line intensity
ratios of the observed transitions at both the center (position A),
and at the position of the molecular cloud complex in the northwestern
arm (position B).  All $\co$ line ratios listed in Table\,\ref{m51rat}
are for a $21''$ beam (roughly corresponding to a circle of 0.5 kpc
radius). The isotopic intensity ratios $\co/\13co$ also refer to that
beam size in the $J$=1--0 and $J$=2--1 transitions, but to a smaller
$14''$ beam in the $J$=3--2 transition. However, the isotopic
intensity ratios appear to vary little in the inner part of M~51
(cf. lit. cited) so that we expect that observation with a larger beam
would have yielded very similar values for the $J$=3--2 transition.
Note that the [CII]/CO(2--1) ratio was determined after convolution of
the $J$=2--1 $\co$ map to the 55$''$ beam of the [CII] observations.
The [CII] intensities (Crawford et al. 1985; Stacey et al. 1991; see
also Negishi et al. 2001) were converted to velocity-integrated
temperatures to obtain the line ratios in Table\,\ref{m51rat}.

The $\co$ and $\13co$ ratios of M~51 are quite different from those of
the other galaxies (NGC~6946, M~83, IC~342, Maffei~2; Israel $\&$ Baas
2001, 2003) we have studied so far.  The ratio of [CI] to $J$=2--1
$\co$ intensities is relatively high, but that of [CII] to $J$=2--1
$\co$ is not dissimilar from those seen in other galaxies. It
resembles, in particular, the ratios seen in galaxies with central
starbursts.  We suspect that this similarity is caused by the large
(radius $R \approx 1.3$ kpc) aperture used to measure the [CII] line
in M~51. We suspect that the very center of M~51, weak in CO and HI,
contributes relatively little, and that most of the emission in the
aperture comes from the inner spiral arms with their strong emission
from star-forming regions.  Bright [CII] emission indicates the
occurrence of both high temperatures and high gas densities as the
critical values for this transition are $T_{\rm kin} \geq$ 91 K and $n
\geq 3500 \cc$. However, such values must be reconciled with the
relatively high CO opacities implied by the modest isotopic intensity
ratios in the lower CO transitions.  In the arm position B, isotopic
ratios increase with increasing $J$ level, but at the center position
A the isotopic ratio is lowest in the $J$=2--1 transition.  This
reflects significant differences in the molecular gas properties at
the two positions sampled.

\section{Analysis}

\subsection{Modelling of CO}

\begin{table*}
\caption[]{Model parameters}
\begin{flushleft}
\begin{tabular}{lccccccc}
\hline
\noalign{\smallskip} 
Model & \multicolumn{3}{c}{Component 1}    	  & \multicolumn{3}{c}{Component 2}  		  & Relative\\
     & Kinetic	    & Gas	 & CO Column   	  & Kinetic 	& Gas 	      	   & CO Column    & $J$=2--1 $\co$ \\
     & Temperature  & Density    & Density     	  & Temperature & Density     	   & Density      & Emission \\
     & $T_{\rm k}$  & $n(\h2$) & $N$(CO)/d$V$	  & $T_{\rm k}$ & $n(\h2$)  	   & $N$(CO)/d$V$ & Component\\
     & (K)     	    & ($\cc$)    & ($\cm2 (\kms)^{-1}$) & (K) 	& ($\cc$) & ($\cm2 (\kms)^{-1}$)  & 1:2 \\
\noalign{\smallskip}
\hline
\noalign{\smallskip}
1 (Center) & 100 &   100 & 1.0$\times10^{17}$ &  60 &   1000 &  3.0$\times10^{17}$ & 0.65 : 0.35 \\
2 (Center) & 150 &  1000 & 0.6$\times10^{17}$ &  20 &   3000 &  3.0$\times10^{17}$ & 0.55 : 0.45 \\
3 (Arm) & 100 &   100 & 1.0$\times10^{17}$ & 150 &   1000 &  3.0$\times10^{17}$ & 0.80 : 0.20 \\
\noalign{\smallskip}
\hline
\end{tabular}
\label{m51model}
\end{flushleft}
\end{table*}

\begin{table*}
\caption[]{Beam-averaged results}
\begin{center}
\begin{tabular}{lccccccc}
\hline
\noalign{\smallskip} 
Model & \multicolumn{3}{c}{Beam-Averaged Column Densities} & \multicolumn{2}{c}{Face-on Mass Density}       & Relative Mass & Ratio $N(\h2)/I({\rm CO})$ \\
   & $N$(CO) & $N$(C)    & $N(\h2)$ & $\sigma(\h2)$ & $\sigma_{\rm gas}^{a}$ & Components 1:2 & $X$ \\
   & \multicolumn{2}{c}{($10^{18} \cm2$)} & ($10^{21} \cm2$) & \multicolumn{2}{c}{($\Msun$/pc$^{-2}$)} & & ($10^{20}$\,mol$\cm2/\kkms$)\\
\noalign{\smallskip}
\hline
\noalign{\smallskip}
\multicolumn{8}{c}{M~51; $N_{\rm H}/N_{\rm C}^{b}$ = 1250; $N$(HI)$^{c} = 2\times 10^{20} \cm2$} \\
\noalign{\smallskip}
\hline
\noalign{\smallskip}
1  & 0.36  & 4.4  & 2.9  & 43  & 60  & 0.6 : 0.4  & 0.74 \\
2  & 0.36  & 1.2  & 1.0  & 15  & 20  & 0.4 : 0.6  & 0.25 \\
3  & 0.42  & 3.3  & 2.2  & 33  & 47  & 0.8 : 0.2  & 0.41 \\
\noalign{\smallskip}
\hline
\end{tabular}
\label{m51beamav}
\end{center}
Note: a. Sum of $\h2$ and HI multiplied by 1.35 to take into account
the contribution by Helium; b. See text, Sect. 4.2; c. HI column
density valid for both Center and (nearby) Arm; Tilanus $\&$ Allen
(1991).
\end{table*}

We have modelled the observed $\co$ and $\13co$ line intensities and
ratios with the large-velocity gradient (LVG) radiative transfer
models described by Jansen (1995) and Jansen et al. (1994). These
provide model line intensities as a function of three input parameters
per molecular gas component: gas kinetic temperature $T_{\rm k}$,
molecular hydrogen density $n(H_{2})$ and the CO column density per
unit velocity $N({\rm CO})$/d$V$. By comparing model to observed line
{\it ratios}, we may identify the physical parameters best describing
the actual conditions at the observed positions.  Beam-averaged
properties are determined by comparing observed and model
intensities. In principle, with seven measured line intensities of two
isotopes, properties of a single gas component are overdetermined as
only five independent observables are required.  We found that no fit
based on a single gas component is capable of matching the data at
either of the observed positions.

However, we obtained good fits based on {\it two} gas components. In
order to reduce the number of free parameters, we assumed identical CO
isotopical abundances for both gas components. In a small number of
starburst galaxy centers (NGC~253, NGC~4945, M~82, IC~342, He~2-10),
values of $40\pm10$ have been suggested for the isotopical abundance
[$^{12}$CO]/[$^{13}$CO] (Mauersberger $\&$ Henkel 1993; Henkel et
al. 1993, 1994, 1998; Bayet et al. 2004), somewhat higher than the
characteristic value of 20--25 for the Milky Way nuclear region
(Wilson \& Rood 1994). Although M~51 does not have a starburst nucleus
but rather a low-luminosity AGN, vigorously starforming spiral arms
already occur at small radii. For this reason we have adopted an
abundance value of [$^{12}$CO]/[$^{13}$CO] = 40 in our models.  We
identified acceptable fits by searching a grid of model parameter
combinations (10 K $\leq T_{\rm k} \leq $ 150 K, $10^{2} \cc \leq
n(\h2) \leq 10^{5} \cc$, $6 \times 10^{15} \cm2 \leq N(CO)/dV \leq 3
\times 10^{18} \cm2$) for model line ratios matching the observed set,
with the relative contribution of the two components as a free
parameter.  Solutions obtained in this way are not unique, but rather
delineate a range of values in distinct regions of parameter space.
For instance, variations in input parameters may to some extent
compensate one another, producing identical line ratios for somewhat
different combinations of input parameters.  Among all possible
solution sets, we have in any case rejected those in which the denser
gas component is also hotter than the more tenuous component, because
we consider the large pressure imbalances implied by such solutions
physically implausible, certainly on the kiloparsec scales observed.

The results for the two positions in M~51 are summarized in
Table\,\ref{m51model}. The physical conditions applying to the
molecular complex in the arm (position B) are very well determined by
the observations.  The emission is wholly dominated by fairly hot gas
with a kinetic temperature of the order of $T_{\rm kin}$ = 100 K.
Low-density gas with $n_{\h2} \approx 100 \cc$ contributes most
($80\pm10\%$) of the observed emission, the remainder ($20\pm10\%$)
coming from gas at a moderately high density $n_{\h2} \approx 1000
\cc$ (fortuitously, the masses of the two components are in almost the
same proportion -- cf. Table\,\ref{m51beamav}).  These parameters
provide a very good fit as is evident from Table\,\ref{m51rat}; no
other set of values does.

In contrast, the physical conditions determining the emission from the
center of M~51 are more difficult to pin down.  Within the errors, two
large sets of conditions yield line ratios consistent with the
observed values.  The first set is very similar to the arm conditions
found above: about two thirds of the emission must come from
relatively hot, fairly tenuous gas ($T_{\rm kin} \approx$ 100 K;
$n_{\h2} \approx 100 \cc$) while the remainder is contributed by
denser and cooler gas ($T_{\rm kin}$ = 40$\pm$20 K; $n_{\h2} \approx
1000 \cc$).  The second set of possible solutions requires all
emission to come from moderately dense ($n_{\h2} = 1000 - 3000 \cc$)
gas.  Half of the emission is then contributed by fairly cold ($T_{\rm
kin} \approx 20$ K) and half by fairly warm ($T_{\rm kin} \approx 150$
K) gas. The present observations do not allow us to distinguish
between these two possibilities; both are equally likely.

\subsection{Beam-averaged molecular gas properties}

The chemical models presented by van Dishoeck $\&$ Black (1988) show a
strong dependence of the $N({\rm C})/N({\rm CO})$ column density ratio
(i.e. how much more carbon there is than the fraction tied up in
carbon monoxide) on the total carbon and molecular hydrogen column
densities. It thus provides a relation between the amounts of carbon
monoxide, neutral carbon and ionized carbon that produce the observed
line intensities from CO, C$^{\circ}$, and C$^{+}$.  In our analysis,
we have assumed that the kinetic temperatures, $\h2$ densities and
filling factors implied by the CO analysis equally apply to the [CI]
and [CII] emission. In principle, one can then solve for column
densities $N({\rm CI})/$d$V$ and $N({\rm CII})$/d$V$. In practice, the
column density of one component is usually well-determined, but that
of the other is more or less degenerate.  For this reason, we solved
for identical velocity dispersions in the two gas components.
Finally, we related total carbon (i.e. C$^{\circ}$ + C$^{+}$ + CO)
column densities to molecular hydrogen column densities by using an
estimated [C]/[H] gas-phase abundance ratio.

Although the analysis in terms of two gas components is rather
superior to that assuming a single component, it is still not fully
realistic.  For instance, the assumption of identical beam-filling
factors or identical velocity dispersions for the various species
($\co, \13co$, C$^{\circ}$, and C$^{+}$) is not a priori plausible.
Fortunately, these assumptions are useful but not critical in the
determination of beam-averaged parameters.  If, by way of example, we
assume a smaller beam-filling factor, the model cloud intensity
increases. This generally implies a higher model column-density which,
however, is more strongly diluted by the beam.  The beam-averaged
column-density is modified only by the degree of nonlinearity in the
response of the model parameters to a change in filling factor, {\it
not} by the magnitude of that change.

Oxygen abundances and gradients for M~51 were taken from Vila-Costas
$\&$ Edmunds 1992 and Zaritzky et al. 1994.  From the results
published by Garnett et al. (1999) we estimate that at the relevant
high metallicities [C]/[O] $\approx$ 1, leading us to adopt a carbon
abundance [C]/[H] = [O]/[H] = $3.0\pm0.5\times10^{-3}$.  As a
significant fraction of carbon is tied up in dust particles and thus
unavailable in the gas-phase, we have adopted a fractional correction
factor $\delta_{\rm c}$ = 0.27 (see for instance van Dishoeck $\&$
Black 1988), so that $N_{\rm H}/N_{\rm C}$ = [2$N(\h2) +
N$(HI)]/[$N$(CO) + $N$(CII) + $N$(CI)] = 1250, uncertain by about a
factor of two. In Table\,\ref{m51beamav} we present beam-averaged column
densities for both CO and C (C$^{\rm o}$ + C$^{+}$), and $\h2$ column
densities derived under the assumptions just discussed, as well as the
face-on mass densities. As the observed peak CO intensities are
significantly below the model peak intensities, only a small fraction
of the (large) beam surface area can be filled with emitting material.
For position A (central region) we find a beam area filling factor of
$70\pm15$ for either model, whereas we find a filling factor of about
20 for position B (the spiral arm CO complex).

In order to gauge the reliability of these results, we have explored
the effect of the assumptions discussed above.  If we do not assume
identical velocity dispersions for the two gas components, but e.g.
dispersions proportional to the kinetic temperature, derived $\h2$
columns and mass-densities change by amounts varying from $10\%$ to
$20\%$ in the three models described.  If we do not assume equal beam
filling factors for CO and C$^{+}$ (which greatly exceeds the
contribution by C$^{\rm o}$, see below), but vary that of C$^{+}$ by
e.g a factor of two, resulting $\h2$ columns and mass-densities change
by amounts varying from $25\%$ to $30\%$. Differences in beam-filling
factors cannot be very much larger, because then the CO and [CII]
intensities, the [C]/[H] and [C]/[CO] abundances, and the $\h2$ column
densities become mutually incompatible.

\subsection{Discussion}

As mentioned above, conditions in the center are fully described by
either of two models that cannot be distinguished as both are fully
consistent with the observed line ratios and intensities.
Observations with much higher resolution using the Smithsonian
Millimeter Array (SMA) recovered only a small fraction of the total
$J$=3--2 $\co$ flux. They showed, however, the existence of a compact
(a few arcsec in size) molecular cloud centered on the nucleus
(Matsushita et al. 2004) not resolved in our maps. For this cloud, the
authors estimated high densities and temperatures that seem more or
less consistent with those of our Model 2/Component 1. However, as
they used only single-component LVG fitting such resemblance may be
fortuitous.  

Our two models have in common that they explain the observed emission
by requiring the existence of at least two gas components at different
densities and temperatures.  Nevertheless, they imply very different
ISM conditions as these kinetic temperatures and densities vary
significantly between models.  The implied hydrogen column densities,
mass surface densities and CO-to-H$_{2}$ conversion factor $X$ also
differ significantly, by a factor of three.  That difference is almost
entirely caused by the fact that at lower kinetic temperatures a much
larger ionized carbon (C$^{+}$) column density is required to produce
the {\it same observed [CII] line flux} than at higher kinetic
temperatures. 

Beam-averaged {\it neutral} carbon to carbon monoxide column density
ratios range from $N$(C$^{\rm o}$)/$N$(CO) = 0.4--1.0, and are very
comparable to those found in other galaxy centers (White et al. 1994;
Israel et al. 1995; Stutzki et al. 1997; Petitpas $\&$ Wilson 1998;
Israel $\&$ Baas, 2001, 2003). 

Our model results agree well with the parameters that we would have
obtained by applying the PDR models presented by Kaufman et al. (1999)
to the observed intensities and ratios of the CO, [CI], [CII] lines,
and the intensity of the far-infrared continuum (Rand et al. 1992).
For the observed values, the PDR models predict the presence of
molecular gas at temperatures $T_{\rm kin} \approx$ 100 K, and at gas
densities 10$^{3} \cc$ illuminated by a fairly modest ambient
radiation field with log $G_{\rm o}$ = 1.0 ($G_{\rm o}$ expressed in
units of the Habing 1968 field: $1.6\times10^{-3}$ erg s$^{-1}$
cm$^{-2}$).

Our models also imply the central regions of M~51 to have a
CO-to-H$_{2}$ conversion factor $X = 0.50\pm0.25 \times 10^{20}$ $\h2$
mol cm$^{-2}$/$\kkms$, i.e. roughly a factor of four lower than the
commonly used Galactic $X$ value for the Solar Neighbourhood.
However, such low values are not uncommon for galactic centers (Israel
$\&$ Baas 2003 and references therein).  For M~51, virtually identical
results have also been obtained independently by Gu\'elin et
al. (1995) from an analysis of the 1.2$\mu$m emission from cold dust,
and by Nakai $\&$ Kuno (1995) when one considers their results for HII
regions with one arcminute from the nucleus.  Thus, we consider that
the CO-to-H$_{2}$ conversion factor $X$ is unusually well-established
in the center of M~51; there can be little doubt that it is indeed
much lower than the `standard' Milky Way value (see also Strong et
al. 2004).  We emphasize that low values for $X$ mean that for any
given CO intensity, beam-averaged molecular hydrogen column densities
$N(\h2)$ and indeed $\h2$ masses are much less than estimated from the
`standard' conversion.  This has another important implication. All
{\it molecular species abundances} derived by relating an
independently determined column density to a molecular hydrogen column
density obtained from the `standard' Galactic conversion factor are in
error and {\it must be revised upwards} by a potentially large factor.

\section{Conclusions}

\begin{enumerate}

\item Observation of the central arcminute of the well-studied spiral
galaxy M~51 in various transitions of $\co$ and $\13co$, and in [CI]
shows that the molecular gas resides mostly in the bright inner spiral
arms.

\item The absence of a circumnuclear disk or torus in images of the
center of M~51 whereas position-velocity maps exhibit a compact
central feature exhibiting a rapid solid-body rotation implies that in
other galaxies the presence of the latter cannot be taken as proof for
the presence of the former.

\item In the center of M~51, a dense component with $n(\h2) \approx
10^{3} \cc$ and $T_{kin} \approx$ 100 K is accompanied by either a
{\it less dense} component with $n(\h2) \approx 10^{2} \cc$ at the
same temperature, or a {\it more dense} component with $n(\h2) \approx
3 \times 10^{3} \cc$ at a much lower temperature of $T_{kin} \approx$
20 K. Only a small fraction of all carbon is in CO. Total carbon
column densities are about 7 times the CO column density. At least in
the center of M~51, most of the carbon appears to be in the form of
C$^{+}$. The observed line ratios are consistent with standard PDR
models.

\item Emission from the CO peak in the northwestern arm likewise
originates in (at least) two different gas components. One is modestly
dense ($n(\h2) \approx 10^{3} \cc$) and the other is relatively
tenuous ($n(\h2) \approx 10^{2} \cc$) but more widespread. Both must
be at elevated temperatures $T_{kin}$ = 100--150 K.

\item The center of M~51 has a face-on gas mass density of $40\pm20
\Msun$\, pc$^{-2}$, and a relatively well-established CO-to-H$_{2}$
conversion factor $X = 0.50\pm0.25 \times 10^{20} \h2$ mol
$\cm2/\kkms$, i.e. a factor of four to five lower than commonly
assumed standard Galactic values, but in line with similar
determinations in other galaxy centers.

\end{enumerate}

\acknowledgements

We are indebted to Ewine van Dishoeck and David Jansen for providing us 
with their detailed radiative transfer models. We thank the JCMT personnel 
for their support and help in obtaining the observations discussed in this 
paper.


\begin{thebibliography}{}
%
\bibitem{} Aalto S., H\"uttemeister S., Scoville N. Z. $\&$ Thaddeus P. 
       1999 \apj 522, 165
\bibitem{} Bayet E., Gerin M., Phillips T.G., \& Contursi A., 2004, \aua 427, 45
\bibitem{} Berkhuijsen E. M., Bajaja E. $\&$ Beck R., 1993 \aua 279, 357
\bibitem{} Crawford M.K., Genzel R., Townes C.H. $\&$ Watson D.M., 1985 
	   \aua 291, 755
\bibitem{} Dumke M., Nieten Ch., Thuma G., Wielebinski R. $\&$ Walsh W., 2001 
           \aua 373, 853
\bibitem{} Garc\'ia-Burillo S., Combes F. \& Gerin M., 1993 \aua 274, 148
\bibitem{} Garc\'ia-Burillo S., Gu\'elin M. \& Cernicharo J. 1993 \aua 274, 123
\bibitem{} Garnett D.R., Shields G.A., Peimbert M.,\, et al. 1999 \apj 513, 168
\bibitem{} Gu\'elin M., Zylka R., Mezger P. G., Haslam C. G. T. \& Kreysa, E.,
        1995 \aua 298, 29
\bibitem{} Habing H.J., 1969 Bull. Astron. Inst. Netherlands, 19, 421
\bibitem{} Henkel C., Mauersberger R., Wiklind T., et al. 1993, \aua 268, L17
\bibitem{} Henkel C., Whiteoak J.B., $\&$ Mauersberger R., 1994 \aua 284, 17
\bibitem{} Henkel C., Chin Y.-N, Mauersberger R. $\&$ Whiteoak J.B., 1998 
        \aua 329, 443
\bibitem{} Israel F.P. $\&$ Baas F., 1999 \aua 351, 10
\bibitem{} Israel F.P. $\&$ Baas F., 2001 \aua 371, 433
\bibitem{} Israel F.P. $\&$ Baas F., 2003 \aua 404, 495
\bibitem{} Israel F.P., White G.J. $\&$ Baas F., 1995, \aua 302, 343
\bibitem{} Jansen D.J., 1995, Ph.D. thesis, University of Leiden (NL)
\bibitem{} Jansen D.J., van Dishoeck E.F. $\&$ Black J.H., 1994, \aua, 282, 605
\bibitem{} Kaufman M.J., Wolfire M.G., Hollenbach D.J. $\&$ Luhman M.L., 1999
	\apj 527, 795
\bibitem{} Lo K. Y., Ball R.,Masson C. R., Phillips T. G., Scott S. $\&$ 
           Woody D. P., 1987 \apjl 317, 63
\bibitem{} Matsushita S., Kohno K., Vila-Vilaro B., Tosaki T. $\&$ Kawabe R., 
           1998 \apj 495, 267
\bibitem{} Matsushita S., Sakamoto K., Kuo C.-Y., et al. 2004, \apjl 616, 55
\bibitem{} Mauersberger R. $\&$ Henkel C., 1993 Rev. Mod. Astron. 6, 69
\bibitem{} Mauersberger R., Henkel C., Walsh W. $\&$ Schulz A., 1999 
           \aua 341, 256
\bibitem{} Nakai N., Kuno N., Handa T. $\&$ Sofue Y., 1994 \pasj 46, 527
\bibitem{} Nakai N., $\&$ Kuno N., 1995 \pasj 47, 761
\bibitem{} Negishi T., Onaka T., Chan K.-W., $\&$ Roellig T., 2001, \aua
           375, 566
\bibitem{} Paglione T.A.D., Wall W. F., Young J.S. et al. 2001, \apj 135, 183
\bibitem{} Petitpas G.R., $\&$ Wilson C.D., 1998 \apj 503, 219
\bibitem{} Rand R.J., 1993 \apj 404, 593
\bibitem{} Rand R.J. $\&$ Kulkarni S.R., 1990\apjl 349, 43
\bibitem{} Rand R.J., Kulkarni S.R. $\&$ Rice W., 1992\apj 390, 66
\bibitem{} Regan M.W., Thornley M.D., Helfer T.T., et al. 2001 \apj 561, 218
\bibitem{} Rickard L.J $\&$ Blitz L., 1985 \apjl 292, L57
\bibitem{} Rickard L. J, Turner B. E., Palmer P., Morris M., $\&$ Zuckerman B.
        1977 \apjl 218, L51 
\bibitem{} Rydbeck G., Hjalmarson \AA. \& Rydbeck O. E. H., 1985 \aua 144, 282
\bibitem{} Sakamoto K., Okumura S.K., Ishizuki S. $\&$ Scoville N.Z., 1999 
	\apjs 124, 403
\bibitem{} Sandage A. $\&$ Tammann G.A., 1975 \apj 196, 313
\bibitem{} Sandage A. $\&$ Tammann G.A., 1987, {\it A Revised Shapley-Ames
	Catalog of Bright Galaxies}, second edition, Carnegie Institution of
	Washington Publication 635 (Washington, D.C.: Carnegie Institution of
	Washington).
\bibitem{} Sandqvist Aa., Elfhag T. $\&$ Lindblad, P. O., 1989 A$\&$A 218, 39
\bibitem{} Scoville N.Z., $\&$, Young J. S. 1983 \apj 265, 148
\bibitem{} Scoville N.Z., Yun M. S., Armus L. $\&$ Ford H. 1998 \apjl 493, 63
\bibitem{} Stacey G.J., Geis N., Genzel R.,\, et al. 1985 \aua 373, 423
\bibitem{} Strong A.W., Moskalenko I.V., Reimer O., Digel S., $\&$ Diehl R.,
`       2004 \aua 422, 47
\bibitem{} Stutzki J., Graf U.U., Honingh C.E.,\, et al. 1997, \apjl 477, 33
\bibitem{} Tilanus R.P.J. $\&$ Allen R.J., 1991 \aua 244, 8
\bibitem{} Tosaki T., Kawabe R. $\&$ Taniguchi Y., 1994, in: Astronomy with 
         Millimeter and Submillimeter Wave Interferometry, IAU Colloquium 
         140, ASP Conf. Ser., Vol. 59, Eds. M. Ishiguro $\&$ J. Welch, 
         p. 353
\bibitem{} Tosaki T., Hasegawa T., Shioya Y., Kuno N. $\&$ Matsushita S., 
         2002 \pasj 54, 209
\bibitem{} Tully R.B., 1974 \apjs 27, 437
\bibitem{} Turner J.L. $\&$ Ho P.T.P., 1994 \apj 421, 122
\bibitem{} van Dishoeck E.F. $\&$ Black J.H., 1988, ApJ 334, 771
\bibitem{} Vila-Costas M.B. $\&$ Edmunds M.G. 1992 \mnras 259, 121
\bibitem{} White G.J., Ellison B., Claude S., Dent W.R.F. $\&$
         Matheson D., 1994 \aua284, L23
\bibitem{} Wielebinski R., Dumke M. $\&$ Nieten Ch., 1999 \aua 347, 643
\bibitem{} Wilson T.L., \& Rood R.T., 1994 \araa 32, 191
\bibitem{} Young J.S., Xie S., Tacconi L., et al. 1995 \apjs 98, 219
\bibitem{} Zaritsky D., Kennicutt R.C. $\&$ Huchra J.P., 1994, \apj 420, 87
%
\end{thebibliography}
\end{document}